\documentclass[12pt]{iopart}

\usepackage{graphicx}
\begin{document}

\title{Long-range memory model of trading activity and volatility}

\author{V. Gontis and B. Kaulakys}

\address{Institute of Theoretical Physics and Astronomy of Vilnius
University, Gostauto 12, Vilnius, LT-01108, Lithuania}
\ead{gontis@itpa.lt}
\begin{abstract}
Earlier we proposed the stochastic point process model, which
reproduces a variety of self-affine time series exhibiting power
spectral density $S(f)$ scaling as power of the frequency $f$ and
derived a stochastic differential equation with the same long
range memory properties. Here we present a stochastic differential
equation as a dynamical model of the observed memory in the
financial time series. The continuous stochastic process
reproduces the statistical properties of the trading activity and
serves as a background model for the modeling waiting time, return
and volatility. Empirically observed statistical properties:
exponents of the power-law probability distributions and power
spectral density of the long-range memory financial variables are
reproduced with the same values of few model parameters.
\end{abstract}

 \noindent{\it Keywords}:
Stochastic processes, Scaling in socio-economic systems, Models of
financial markets

\maketitle

\section{Introduction}
Stochastic volatility models of the financial time series are
fundamental to investment, option pricing and risk management
\cite{Engle,Mantegna}. The volatility serves as a quantitative
price diffusion measure in widely accepted stochastic
multiplicative process known as geometric Brownian motion (GBM).
Extensive empirical data analysis of big price movements in the
financial markets confirms the assumption that volatility itself
is a stochastic variable or more generally the function of a
stochastic variable \cite{Fouque}. By analogy with physics we can
assume that speculative prices $p(t)$ change in a "random medium"
described by the random diffusion coefficient. Such analogy may be
reversible from the point that the complex models of stochastic
price movements can be applicable for the description of complex
physical systems such as stochastic resonance, noise induced phase
transitions and high energy physics applications. This analogy
contributes to further development of statistical mechanics -
nonextensive one and superstatistics have been introduced
\cite{Tsallis,Beck}.

Additive-multiplicative stochastic models of the financial
mean-reverting processes provide rich spectrum of shapes for the
probability distribution function (PDF) depending on the model
parameters \cite{Anteneodo}. Such stochastic processes model the
empirical PDF's of volatility, volume and price returns with
success when the appropriate fitting parameters are selected.
Nevertheless, there is the necessity to select the most
appropriate stochastic models able to describe volatility as well
as other variables under the dynamical aspects and the long range
correlation aspects. There is empirical evidence that trading
activity, trading volume, and volatility are stochastic variables
with the long-range correlation \cite{Engle,Plerou,Gabaix} and
this key aspect is not accounted for in widespread models.
Moreover, rather often there is evidence that the models proposed
are characterized only by the short-range time memory
\cite{BookDacorogna}.

Phenomenological descriptions of volatility, known as
heteroscedasticity, have proven to be of extreme importance in the
option price modeling \cite{Engle2}. Autoregressive conditional
heteroscedasticity (ARCH) processes and more sophisticated
structures GARCH are proposed as linear dependencies on previous
values of squared returns and variances \cite{Engle2,Bollerslev}.
These models based on empirically fitted parameters fail in
reproducing power law behavior of the volatility autocorrelation
function. We do believe that the stochastic models with the
limited number of parameters and minimum stochastic variables are
possible and would better reflect the market dynamics and its
response to the external noise.

Recently we investigated analytically and numerically the
properties of stochastic multiplicative point processes
\cite{KaulakysPRE}, derived formula for the power spectrum and
related the model with the general form of multiplicative
stochastic differential equation \cite{KaulakysPhA}. Preliminary
the comparison of the model with the empirical data of spectrum
and probability distribution of stock market trading activity
\cite{GontisPhA} stimulated us to work on the definition of more
detailed model. The extensive empirical analysis of the financial
market data, supporting the idea that the long-range volatility
correlations arise from trading activity, provides valuable
background for further development of the long-ranged memory
stochastic models \cite{Plerou,Gabaix}. We will present the
stochastic model of trading activity with the long-range
correlation and will investigate its' connection to the stochastic
modeling of volatility and returns.

\section{Stochastic model of interevent time}

Earlier we proposed the stochastic point process model, which
reproduced a variety of self-affine time series exhibiting the
power spectral density $S(f)$ scaling as power of the frequency
$f$ \cite{KaulakysPRE,GontisPhA}. The time interval between point
events in this model fluctuates as a stochastic variable described
by the multiplicative iteration equation
\begin{equation}
\tau_{k+1}=\tau_{k}+\gamma\tau_{k}^{2\mu-1}+\sigma\tau_{k}^{\mu}\varepsilon_{k}.
\label{eq:tauiterat}
\end{equation}
Here interevent time $\tau_{k}=t_{k+1}-t_{k}$ between subsequent
events $k$ and $k+1$ fluctuates due to the random perturbation by
a sequence of uncorrelated normally distributed random variable
$\{\varepsilon_{k}\}$ with the zero expectation and unit variance,
$\sigma$ denotes the standard deviation of the white noise and
$\gamma\ll 1$ is a coefficient of the nonlinear damping. It has
been shown analytically and numerically
\cite{KaulakysPRE,GontisPhA} that the point process with
stochastic interevent time \eref{eq:tauiterat} may generate
signals with the power-law distributions of the signal intensity
and $1/f^{\beta}$ noise. The corresponding Ito stochastic
differential equation for the variable $\tau(t)$ as a function of
the actual time can be written as
\begin{equation}
d\tau=\gamma\tau^{2\mu-2}\rmd t+\sigma\tau^{\mu-1/2}\rmd W,
\label{eq:taustoch}
\end{equation}
where $W$ is a standard random Wiener process. Eq.
\eref{eq:taustoch} describes the continuous stochastic variable
$\tau(t)$ which can be assumed as slowly diffusing mean interevent
time of Poisson process with the stochastic rate $1/\tau(t)$. We
put the modulated Poisson process  into the background of the
long-range memory point process model.

The diffusion of $\tau$ must be restricted at least from the side
of high values. Therefore we introduce a new term
$-\frac{m}{2}\sigma^2\left(\frac{\tau}{\tau_{0}}\right)^m\tau^{2\mu-2}$
into the Eq. \eref{eq:taustoch}, which produces the exponential
diffusion reversion in equation
\begin{equation}
d\tau=\left[\gamma-\frac{m}{2}\sigma^2\left(\frac{\tau}{\tau_{0}}\right)^m\right]\tau^{2\mu-2}\rmd
t+\sigma\tau^{\mu-1/2}\rmd W, \label{eq:taustoch2}
\end{equation}
where $m$ and $\tau_{0}$ are the power and value of the diffusion
reversion, respectively. The associated Fokker-Plank equation with
the zero flow will give the simple stationary PDF
\begin{equation}
P(\tau)\sim\tau^{\alpha+1}\exp\left[-\left(\frac{\tau}{\tau_{0}}\right)^m\right]\label{eq:taudistrib}
\end{equation}
with  $\alpha=2(\gamma_{\sigma}-\mu)$, where
$\gamma_{\sigma}=\gamma/\sigma^2$. We define the conditional
probability of interevent time $\tau_{\mathrm{p}}$ in the
modulated Poisson point process with stochastic rate $1/\tau$ as
\begin{equation}
\varphi(\tau_{\mathrm{p}}|\tau)=\frac{1}{\tau}\exp\left[-\frac{\tau_{\mathrm{p}}}{\tau}\right].\label{eq:taupoisson}
\end{equation}
Then the long time distribution $\varphi(\tau_{\mathrm{p}})$ of
interevent time $\tau_{\mathrm{p}}$ has the integral form
\begin{equation}
\varphi(\tau_{\mathrm{p}})=C\int_{0}^{\infty}\exp\left[-\frac{\tau_{\mathrm{p}}}{\tau}\right]\tau^{\alpha}\exp\left[-\left(\frac{\tau}{\tau_{0}}\right)^m\right]\rmd
\tau,\label{eq:taupdistrib}
\end{equation}
with $C$ defined from the normalization,
$\int_{0}^{\infty}\varphi(\tau_{\mathrm{p}})\rmd
\tau_{\mathrm{p}}=1$. In the case of pure exponential diffusion
reversion, $m=1$, PDF \eref{eq:taupdistrib} has a simple form
\begin{equation}
\varphi(\tau_{\mathrm{p}})=\frac{2}{\Gamma(2+\alpha)\tau_{0}}\left(\frac{\tau_{\mathrm{p}}}{\tau_{0}}\right)
^{\frac{1+\alpha}{2}}\mathrm{K_{(1+\alpha)}}\left(2
\sqrt{\frac{\tau_{\mathrm{p}}}{\tau_{0}}}\right),\label{eq:taupintegr}
\end{equation}
where $\mathrm{K_{\alpha}}\left(z\right)$ denotes the modified
Bessel function of the second kind. For $m>1$ more complicated
structures of  distribution $\varphi(\tau_{\mathrm{p}})$ expressed
in terms of hypergeometric functions arise.

\section{Stochastic model of flow of points or events }

The introduced stochastic multiplicative model of interevent time,
the interval between trades in the financial market, defines the
model of event flow $n$.  First of all we apply Ito transformation
of variables introducing flow of events $n(t)=\frac{1}{\tau(t)}$.
The stochastic differential equation for $n$ follows from Eq.
\eref{eq:taustoch},
\begin{equation}
\rmd
n=\sigma^2\left[(1-\gamma_{\sigma})+\frac{m}{2}\left(\frac{n_{0}}{n}\right)^{m}\right]n^{2\eta-1}\rmd
t+\sigma n^{\eta}\rmd W, \label{eq:nstoch}
\end{equation}
where $\eta=\frac{5}{2}-\mu$ and $n_{0}=1/\tau_{0}$. Eq.
\eref{eq:nstoch} describes stochastic process $n$ with PDF
\begin{equation}
P(n)\sim\frac{1}{n^{\lambda}}\exp\left\{-\left(\frac{n_{\mathrm{0}}}{n}\right)^m
\right\},\quad\lambda=2(\eta-1+\gamma_{\sigma}),\label{eq:ndistr}
\end{equation}
and power spectrum $S(f)$ \cite{KaulakysPRE,KaulakysPhA,GontisPhA}
\begin{equation}
S(f)\sim\frac{1}{f^{\beta}},\quad\beta=2-\frac{3-2\gamma_{\sigma}}{2\eta-2}.
\label{eq:nspekt}
\end{equation}
Noteworthy, that in the proposed model  only two parameters,
$\gamma_{\sigma}$ and $\eta$ (or $\mu$), define exponents
$\lambda$ and $\beta$  of two power-law statistics, i.e. of PDF
and power spectrum. Time scaling parameter $\sigma^2$ in Eq.
\eref{eq:nstoch} can be omitted adjusting the time scale.

Stochastic variable $n$ denotes the number of  events per unit
time interval. One has to integrate the stochastic signal Eq.
\eref{eq:nstoch} in the time interval $\tau_{\mathrm{d}}$ to get
number of events in the selected time window. In this paper we
will denote the integrated number of points or events as
\begin{equation}
N(t,\tau_{\mathrm{d}})=\int_{t}^{t+\tau_{\mathrm{d}}}n(t^{\prime})\rmd
t^{\prime}
\end{equation}
and will call it trading activity in the case of the financial
market. Flow of points or events arises in different fields, such
as physics, economics, cosmology, ecology, neurology, the
Internet, seismology, i.e., electrons, photons, cars, pulses,
events, and so on, or subsequent actions, like seismic events,
neural action potentials, transactions in the financial markets,
human heart beats, biological ion-channel openings, burst errors
in many communication systems, the Internet network packets, etc.
We will discuss possible application of the proposed stochastic
model  to model the trading activity in the financial markets.

\section{Stochastic model of trading activity}

It is widely accepted that in high-frequency financial data not
only the returns but also the waiting times between the
consecutive trades are random variables \cite{Scalas}. Waiting
times between trades do not follow the exponential distribution
and the related point process is not the Poisson one. The
extensive empirical analysis provides evidence that the related
stochastic variable trading activity defined as flow of trades is
stochastic variable with the long range memory \cite{Plerou2}. We
will investigate how the proposed modulated Poisson stochastic
point process  can be adjusted to model trading activity with the
empirically defined statistical properties. Detrended fluctuation
analysis \cite{Plerou2} is one of the methods to define the second
order statistics, the autocorrelation of trading activity. The
histogram of the detrended fluctuation analysis exponents $\nu$
obtained by fits for each of the 1000 US stocks shows a relatively
narrow spread of $\nu$ around the mean value $\nu=0.85\pm0.01$
\cite{Plerou2}. We use relation between the exponents of detrended
fluctuation analysis and the exponents of power spectrum
$\beta=2\nu-1$ \cite{Beran} and in this way define the empirical
value of the exponent for the power spectral density $\beta=0.7$.
Our analysis of the Lithuanian stock exchange data confirmed that
the power spectrum of trading activity is the same for various
liquid stocks even for the emerging markets \cite{Gontis2}. The
histogram of exponents obtained by fits to the cumulative
distributions of trading activites of 1000 US stocks
\cite{Plerou2} gives the value of exponent $\lambda=4.4\pm0.05$
describing the power-law behavior of the trading activity.
Empirical values of $\beta=0.7$ and $\lambda=4.4$ confirm that the
time series of the trading activity in real markets are fractal
with the power law statistics. Time series generated by stochastic
process \eref{eq:nstoch} are fractal in the same sense.

Nevertheless, we face serious complications trying to adjust model
parameters to the empirical data of the financial markets. For the
pure multiplicative model, when $\mu=1$ or $\eta=3/2$, we have to
take $\gamma_{\sigma}=0.85$ to get $\beta=0.7$ and
$\gamma_{\sigma}=1.7$ to get $\lambda=4.4$, i.e. it is impossible
to reproduce the empirical PDF and power spectrum with the same
relaxation parameter $\gamma_{\sigma}$ and exponent of
multiplicativity $\mu$. We have proposed possible solution of this
problem in our previous publications \cite{GontisPhA,Gontis2}
deriving PDF for the trading activity $N$
\begin{equation}
P(N)\sim\left\{
\begin{array}{ll}
\frac{1}{N^{3+\alpha}},& N\ll \gamma^{-1}, \\
\frac{1}{N^{5+2\alpha}},& N\gg \gamma^{-1}.
\end{array}
\right. \label{eq:Ndistrib}
\end{equation}

When $N\gg \gamma^{-1}$ this yields exactly the required value of
$\lambda=5+2\alpha=4.4$ and $\beta=0.7$ for
$\gamma_{\sigma}=0.85$.

Nevertheless, we cannot accept this as the sufficiently accurate
model of the trading activity as the empirical power law
distribution is achieved only for very high values of the trading
activity. Probably this reveals the mechanism how the power law
distribution converges to normal distribution through the growing
values of the exponent, but empirically observed power law
distribution in wide area of $N$ values cannot be reproduced. Let
us notice here that the desirable power law distribution of the
trading activity with the exponent $\lambda=4.4$ may be generated
by the model \eref{eq:nstoch} with  $\eta=5/2$ and
$\gamma_{\sigma}=0.7$. Moreover, only the smallest values of
$\tau$ or high values of $n$ contribute to the power spectral
density of trading activity \cite{KaulakysPhA}. This suggests us
to combine the point process with two values of $\mu$: (i)
$\mu\simeq0$ for the main area of diffusing $\tau$ and $n$ and
(ii) $\mu=1$ for the lowest values of $\tau$ or highest values of
$n$. Therefore, we introduce a new stochastic differential
equation for $n$ combining two powers of multiplicative noise,

\begin{equation}
\rmd
n=\sigma^2\left[(1-\gamma_{\sigma})+\frac{m}{2}\left(\frac{n_{0}}{n}\right)^{m}\right]\frac{n^4}{(n\epsilon+1)^2}\rmd
t+\frac{\sigma n^{5/2}}{(n\epsilon+1)}\rmd W, \label{eq:nstoch2}
\end{equation}
where a new parameter $\epsilon$ defines crossover between two
areas of $n$ diffusion. The corresponding iterative equation of
form \eref{eq:tauiterat} for $\tau_{k}$ in such a case is

\begin{equation}
\tau_{k+1}=\tau_{k}+\left[\gamma-\frac{m}{2}\sigma^2\left(\frac{\tau}{\tau_{0}}\right)^m\right]\frac{\tau_{k}}{(\epsilon+\tau_{k})^2}+\sigma\frac{\tau_{k}}{\epsilon+\tau_{k}}\varepsilon_{k}.
\label{eq:tauiterat2}
\end{equation}

\begin{figure*}
\begin{center}
\includegraphics[width=.8\textwidth]{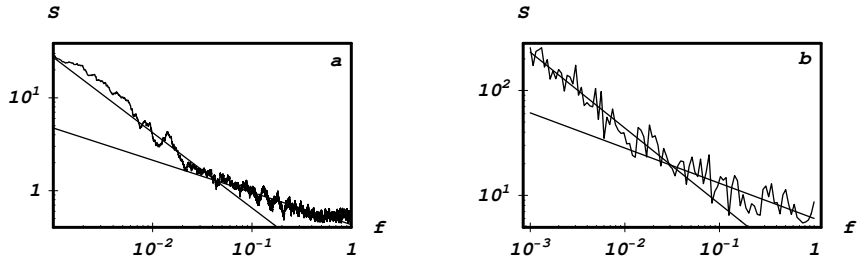}
\hspace{-10pt}
\end{center}
\par
\vspace{-10pt} \caption{Power spectral density $S(f)$ calculated
with parameters $\gamma=0.0004$; $\sigma=0.025$; $\epsilon=0.07$;
$\tau_{0}=1$; $m=6$. Straight lines approximate power spectrum
$S\sim1/f^{\beta_{1,2}}$ with $\beta_{1}=0.33$ and
$\beta_{2}=0.72$: a) $S(f)$ calculated by the Fast Fourier
Transform of $n$ series generated by Eq. \eref{eq:nstoch2},  b)
$S(f)$ averaged over 20 series of 100000 iterations of the flow
$I(t)=\sum_{k}\delta(t-t_{k})$ with the interevent time
$\tau_{k}=t_{k+1}-t_{k}$ generated by Eq. \eref{eq:tauiterat2}.}
\label{fig:1}
\end{figure*}

Eqs. \eref{eq:nstoch2} and \eref{eq:tauiterat2} define related
stochastic variables $n=\frac{1}{\tau}$ and $\tau$, respectively,
and they should reproduce the long range statistical properties of
the trading activity and of waiting time in the financial markets.
We verify this by the numerical calculations. In
figure~\ref{fig:1} we present the power spectral density
calculated for the equivalent processes \eref{eq:nstoch2} and
\eref{eq:tauiterat2} (see \cite{GontisPhA} for details of
calculations). This approach reveals the structure of the power
spectral density in wide range of frequencies and shows that the
model exhibits not one but rather two separate power laws with the
exponents $\beta_{1}=0.33$ and $\beta_{2}=0.72$. From many
numerical calculations performed with the multiplicative point
processes we can conclude that combination of two power laws of
spectral density arise only when multiplicative noise is a
crossover of two power laws, see \eref{eq:nstoch2} and
\eref{eq:tauiterat2}. We will show in the next section that this
may serve as an explanation of two exponents of the power spectrum
in the empirical data of volatility for \verb"S&P 500" companies
\cite{Liu}.

\begin{figure*}
\begin{center}
\includegraphics[width=.8\textwidth]{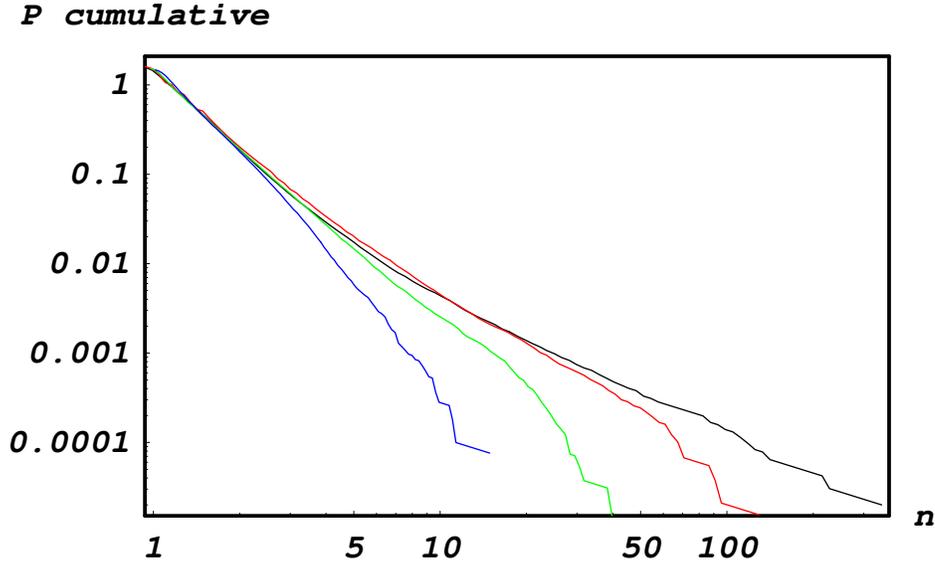}
\hspace{-10pt}
\end{center}
\par
\vspace{-10pt} \caption{Cumulative PDF $P_{>}(n)$ calculated from
the histogram of $N/\tau_{\mathrm{d}}$ generated by Eq.
\eref{eq:tauiterat2} with the increasing time interval
$\tau_{\mathrm{d}}$ from the above curve: $\tau_{\mathrm{d}}=1,
10, 50$ and $250$. Other parameters are as in figure~\ref{fig:1}.}
\label{fig:2}
\end{figure*}

Empirical data of the trading activity statistics must be modeled
by the integrated flow of event $N$ defined in the time interval
$\tau_{\mathrm{d}}\gg\tau_{0}$. In figure~\ref{fig:2} we
demonstrate the cumulative probability distribution functions
  $P_{>}(n)$ calculated from the histogram of $N/\tau_{\mathrm{d}}$ generated by Eq.
\eref{eq:tauiterat2} with increasing time interval
$\tau_{\mathrm{d}}$. This illustrates how  distribution of the
integrated signal $N$ converges to the normal distribution (the
central limit theorem) through growing exponent of the power-law
distribution and provides an evidence that the empirically
observed exponent $\lambda=4.4$ of the power-law distribution of
$N$ \cite{Plerou,Gabaix} can be explained by the proposed model
with the same parameters suitable for description of the power
spectrum of the trading activity.

\begin{figure*}
\begin{center}
\includegraphics[width=.8\textwidth]{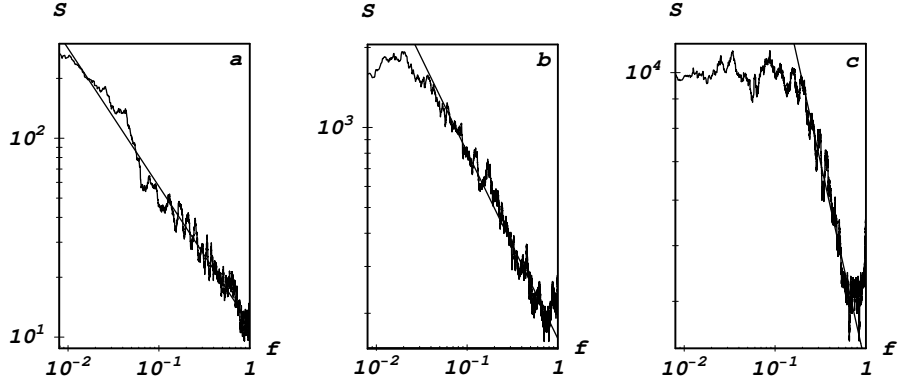}
\hspace{-10pt}
\end{center}
\par
\vspace{-10pt} \caption{Power spectral density of the trading
activity $N$ calculated by the Fast Fourier Transform of $N$
series generated with Eq. \eref{eq:nstoch2} for the same
parameters as in figures~\ref{fig:1} and \ref{fig:2}: a)
$\tau_{\mathrm{d}}=10$; b) $\tau_{\mathrm{d}}=50$; c)
$\tau_{\mathrm{d}}=250$. Straight lines approximate power spectrum
$S\sim 1/f^\beta$, with $\beta=0.7$.} \label{fig:3}
\end{figure*}

The power spectrum of the trading activity $N$ can be calculated
by the Fast Fourier Transform of the generated numerical series.
As illustrated in figure~\ref{fig:3}, the exponents $\beta=0.7$ of
the power spectrum  are independent of $\tau_{\mathrm{d}}$ and
reproduce the empirical results of the detrended fluctuation
analysis \cite{Plerou,Gabaix}.

The same numerical results can be reproduced by continuous
stochastic differential equation \eref{eq:nstoch2} or iteration
equation \eref{eq:tauiterat2}. One can consider the discrete
iterative equation for the interevent time $\tau_{k}$
\eref{eq:tauiterat2} as a method to solve numerically continuous
equation

\begin{equation}
\rmd\tau=\left[\gamma-\frac{m}{2}\sigma^2\left(\frac{\tau}{\tau_{0}}\right)^m\right]\frac{1}{(\epsilon+\tau)^2}\rmd
t+\sigma\frac{\sqrt{\tau}}{\epsilon+\tau}\rmd W.
\label{eq:taucontinuous}
\end{equation}
The continuous equation \eref{eq:nstoch2} follows from the Eq.
\eref{eq:taucontinuous} after change of variables $n=1/\tau$.

We can conclude that the long range memory properties of the
trading activity in the financial markets as well as the PDF can
be modeled by the continuous stochastic differential equation
\eref{eq:nstoch2}. In this model the exponents of the power
spectral density, $\beta$, and of PDF, $\lambda$, are defined by
one parameter $\gamma_{\sigma}=\gamma/\sigma^{2}$. We consider the
continuous equation of the mean interevent time $\tau$ as a model
of slowly varying stochastic rate $1/\tau$ in the modulated
Poisson process

\begin{equation}
\varphi(\tau_{\mathrm{p}}|\tau)=\frac{1}{\tau}\exp\left[-\frac{\tau_{\mathrm{p}}}{\tau}\right].
\label{eq:taupoisson}
\end{equation}

\begin{figure*}
\begin{center}
\includegraphics[width=.8\textwidth]{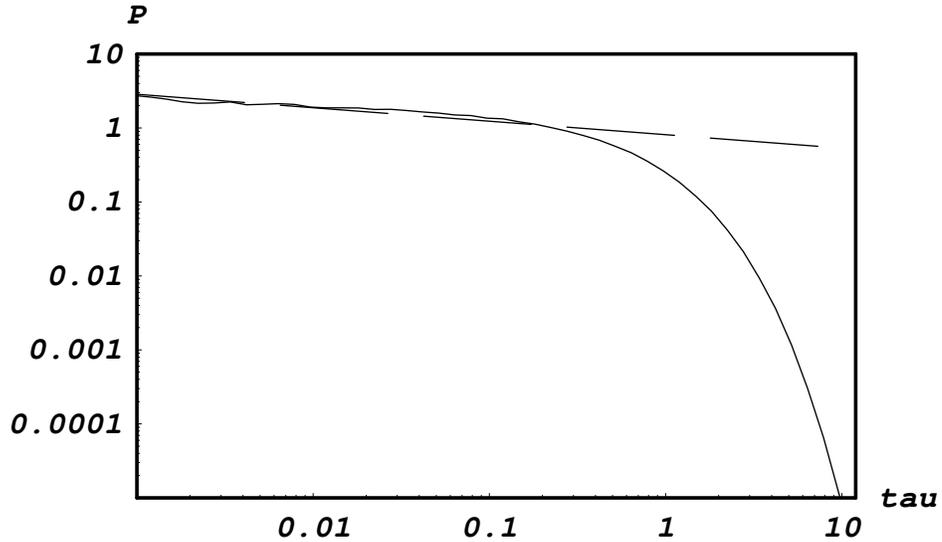}
\hspace{-10pt}
\end{center}
\par
\vspace{-10pt} \caption{Probability distribution function
$P(\tau_{\mathrm{p}})$ calculated from the histogram of
$\tau_{\mathrm{p}}$ generated by Eq. \eref{eq:taupoisson} with
rate calculated from Eq. \eref{eq:taucontinuous}. Used parameters
are $\gamma=0.0004$; $\sigma=0.025$; $\epsilon=0.07$; $\tau_{0}=1$
and $m=6$. Dashed line approximates power law
$P(\tau_{\mathrm{p}})\sim\tau_{\mathrm{p}}^{-0.15}$. }
\label{fig:4}
\end{figure*}

In figure~\ref{fig:4} we demonstrate the probability distribution
functions $P(\tau_{\mathrm{p}})$ calculated from the histogram of
$\tau_{\mathrm{p}}$ generated by Eq. \eref{eq:taupoisson} with the
diffusing mean interevent time calculated from Eq.
\eref{eq:taucontinuous}.

Numerical results show good qualitative agreement with the
empirical data of interevent time probability distribution
measured from few years series of U.S. stock data \cite{Ivanov}.
This enables us to conclude that the proposed stochastic model
captures the main statistical properties including PDF and the
long range correlation of the trading activity in the financial
markets. Furthermore, in the next section we will show  that this
may serve as a background statistical model responsible for the
statistics of return volatility in widely accepted GBM of the
financial asset prices.

\section{Modeling returns and volatility}

We follow an approach developed in \cite{Plerou,Gabaix,Plerou2} to
analyze the empirical data of price fluctuations driven by the
market activity. The basic quantities studied for the individual
stocks are price $p(t)$ and return

\begin{equation}
x(t,\tau_{\mathrm{d}})=\ln p(t+\tau_{\mathrm{d}})-\ln p(t)
\label{eq:return}
\end{equation}

Return $x(t,\tau_{\mathrm{d}})$ over a time interval
$\tau_{\mathrm{d}}$ can be expressed through the subsequent
changes $\delta x_{i}$ due to the trades
$i=1,2....N(t,\tau_{\mathrm{d}})$ in the time interval
$[t,t+\tau_{\mathrm{d}}]$,

\begin{equation}
x(t,\tau_{\mathrm{d}})=\sum_{i=1}^{N(t,\tau_{\mathrm{d}})}\delta
x_{i}. \label{eq:return2}
\end{equation}

We denote the variance of $\delta x_{i}$ calculated over the time
interval $\tau_{\mathrm{d}}$ as $W^{2}(t,\tau_{\mathrm{d}})$. If
$\delta x_{i}$ are mutually independent one can apply the central
limit theorem to sum \eref{eq:return2}. This implies that for the
fixed variance $W^{2}(t,\tau_{\mathrm{d}})$ return
$x(t,\tau_{\mathrm{d}})$ is a normally distributed random variable
with the variance
$W^{2}(t,\tau_{\mathrm{d}})N(t,\tau_{\mathrm{d}})$

\begin{equation}
x(t,\tau_{\mathrm{d}})=W(t,\tau_{\mathrm{d}})\sqrt{N(t,\tau_{\mathrm{d}})}\varepsilon_{t}
. \label{eq:return3}
\end{equation}

\begin{figure*}
\begin{center}
\includegraphics[width=.6\textwidth]{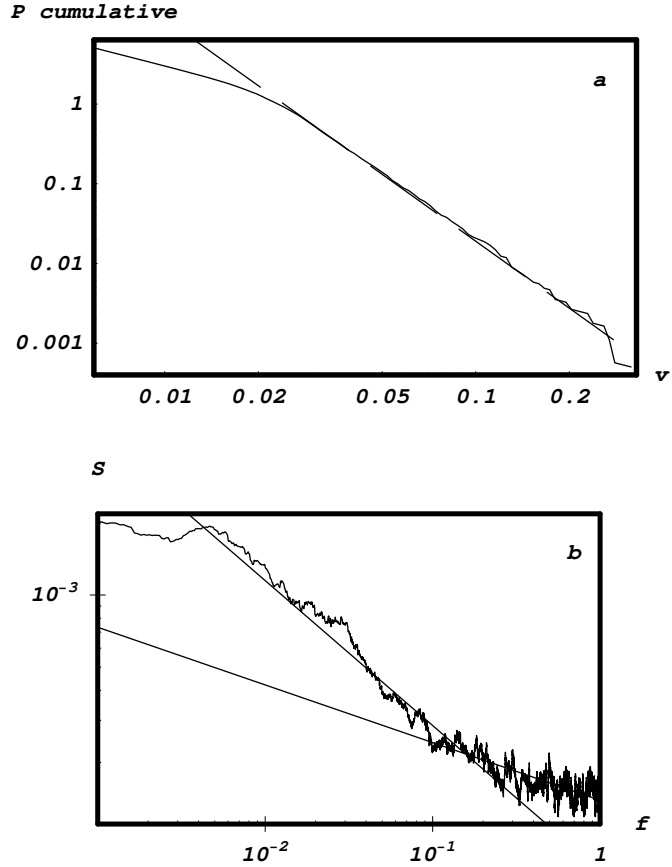}
\hspace{-10pt}
\end{center}
\par
\vspace{-10pt} \caption{ (a) Cumulative probability distribution
function of the volatility, $P_{>}(\overline{v})$, averaged over
10 intervals calculated from the series of $n(t)$ generated by
Eqs. \eref{eq:nstoch2} and \eref{eq:return4}, all parameters are
the same as in previous calculations.  Dashed line approximates
the power law $P(\overline{v})\sim 1/\overline{v}^{2.8}$. (b)
Power spectral density $S(f)$ of $v$ calculated from FFT of the
same series $n(t)$. Straight lines approximate power spectral
density $S\sim 1/f^{\beta_{1,2}}$ with $\beta_{1}=0.6$ and
$\beta_{2}=0.24$.} \label{fig:5}
\end{figure*}

Empirical test of conditional probability
$P(x(t,\tau_{\mathrm{d}})|W(t,\tau_{\mathrm{d}}))$ \cite{Plerou}
confirms its' Gaussian form, and the unconditional distribution
$P(x(t,\tau_{\mathrm{d}}))$ is a power-law with the cumulative
exponent $3$. This implies that the power-law tails of returns are
largely due to those of $W(t,\tau_{\mathrm{d}})$. Here we refer to
the theory of price diffusion as a mechanistic random process
\cite{Farmer,Farmer2}. For this idealized model the short term
price diffusion depends on the limit order removal and this way is
related to the market order flow. Furthermore, the empirical
analysis confirms that the volatility calculated for the fixed
number of transactions has the long memory properties as well and
it is correlated with real time volatility \cite{Farmer3}. We
accumulate all these results into strong assumption that standard
deviation $W(t,\tau_{\mathrm{d}})$ may be proportional to the
square root of the trading activity, i.e.,
$W(t,\tau_{\mathrm{d}})=k \sqrt{N(t,\tau_{\mathrm{d}})}$. This
enables us to propose a very simple model of return

\begin{equation}
x(t,\tau_{\mathrm{d}})=k N(t,\tau_{\mathrm{d}})\varepsilon_{t}
\label{eq:return4}
\end{equation}
and related model of volatility $v=|x(t,\tau_{\mathrm{d}})|$ based
on the proposed model of trading activity \eref{eq:nstoch2}. We
generate series of trade flow $n(t)$ numerically solving Eq.
\eref{eq:nstoch2} with variable steps of time $\Delta
t_{i}=h_{i}=n_{0}/n_{i}$ and calculate the trading activity in
subsequent time intervals $\tau_{\mathrm{d}}$ as
$N(t,\tau_{\mathrm{d}})=\int_{t}^{t+\tau_{\mathrm{d}}}n(t^{\prime})dt^{\prime}$.
This enables us to generate series of return
$x(t,\tau_{\mathrm{d}})$, of volatility
$v(t,\tau_{\mathrm{d}})=|x(t,\tau_{\mathrm{d}})|$ and of the
averaged volatility
$\overline{v}=\frac{1}{m}\sum_{i=1}^{i=m}v(t_{i},\tau_{\mathrm{d}})$.

In figure~\ref{fig:5} we demonstrate cumulative distribution of
$\overline{v}$ and the power spectral density of
$v(t,\tau_{\mathrm{d}})$ calculated from FFT. We see that proposed
model enables us to catch up the main features of the volatility:
the power law distribution with exponent $2.8$ and power spectral
density with two exponents $\beta_{1}=0.6$ and $\beta_{2}=0.24$.
This is in a good agreement with the empirical data
\cite{Liu,Farmer3}.

\section{Conclusions}

Earlier proposed stochastic point process model
\cite{GontisPhA,Gontis2} as a possible model of trading activity
in the financial markets has to be elaborated. First of all, we
define that the long-range memory fluctuations of trading activity
in financial markets may be considered as background stochastic
process responsible for the fractal properties of other financial
variables. Waiting time in the sequence of trades more likely is
double stochastic process, i.e., Poisson process with the
stochastic rate defined as a stand-alone stochastic variable. We
consider the stochastic rate  as continuous one and model it by
the stochastic differential equation, exhibiting long-range memory
properties. We reconsider previous stochastic point process  as
continuous process and propose the related nonlinear stochastic
differential equation with the same statistical properties
\cite{KaulakysPhA}. One more elaboration of the model is needed to
build up the stochastic process with the statistical properties
similar to the empirically defined properties of trading activity
in the financial markets. We combine the market response function
to the noise as consisting of two different powers: one
responsible for the probability distribution function and another
responsible for the power spectral density. The proposed new form
of the continuous stochastic differential equation enables us to
reproduce the main statistical properties of the trading activity
and waiting time, observed in the financial markets. More precise
model definition enables us to reproduce power spectral density
with two different scaling exponents. This provides an evidence
that the market behavior is dependant on the level of activity and
two stages: calm and excited must be considered. We proposed a
very simple model to reproduce the statistical properties of
return and volatility. More sophisticated approach has to be
elaborated to account for the leverage effect and other specific
features of the market.

\section*{Acknowledgements}
We acknowledge support by the Lithuanian State Science and Studies
Foundation.

\end{document}